\documentclass[11pt,a4paper]{article}
\pdfoutput=1

\usepackage{jcappubtweaked}
\usepackage{amsmath,amssymb,amsfonts,mathrsfs,mathdots}

\renewcommand{\d}{\mathrm{d}}
\newcommand{\pder}[2][{}]{\frac{\partial {#1}}{\partial {#2}}}

\renewcommand{\O}{\mathcal{O}}
\newcommand{\cor}[2][u]{\langle {#2}\rangle_{#1}}
\newcommand{\bvec}[1]{\boldsymbol{#1}}

\subheader{\hfill \rm{MAD-TH-12-08}}

\title{
Consistency condition for inflation from (broken) conformal symmetry
}

\author[a]{Koenraad Schalm}
\author[b,c]{Gary Shiu}
\author[a]{Ted van der Aalst}
\affiliation[a]{Instituut-Lorentz for Theoretical Physics, Universiteit Leiden,\\
\mbox{Niels Bohrweg 2, Leiden, {The Netherlands}}}
\affiliation[b]{Department of Physics, University of Wisconsin-Madison,\\
\mbox{Madison, WI 53706, USA}}
\affiliation[c]{Department of Physics and Institute for Advanced Study, \\
Hong Kong University of Science and Technology, Hong Kong}
\emailAdd{kschalm@lorentz.leidenuniv.nl}
\emailAdd{shiu@physics.wisc.edu}
\emailAdd{vdaalst@lorentz.leidenuniv.nl}

\abstract{
We investigate the symmetry constraints on the bispectrum, i.e. the three-point correlation
function of primordial density fluctuations, in slow-roll inflation.
It follows from the defining property of slow-roll inflation that primordial correlation functions inherit most of their structure from weakly broken de Sitter symmetries.
Using holographic techniques borrowed from the AdS/CFT correspondence, the symmetry constraints on the bispectrum can be mapped to a set of stress-tensor Ward identities in a weakly broken $2+1$-dimensional Euclidean CFT.
We construct the consistency condition  from these Ward identities using conformal perturbation theory.
This requires a second order Ward identity and the use of the evolution equation. Our result also illustrates a subtle difference between conformal perturbation theory and the slow-roll expansion.
}


\notoc
\begin{document}
\maketitle
\newpage
\section{Introduction}
In recent years,
there has been much effort in the calculation and understanding of
the power spectrum, bispectrum \cite{Maldacena0210603,Acquaviva0209156,Seery0503692,Chen0605045}
and trispectrum
\cite{Huang0610235,Arroja08021167,Seery0610210,Arroja09053641,Chen09053494}
of
primordial
curvature perturbations generated in
an
inflationary universe.
These correlation functions provide more and more refined maps between theory and observation.
 While current data so far only allows us to extract definitively
 the two-point correlation, it is widely believed that primordial non-Gaussianities, if detected, would open up a novel route to probe high scale physics \cite{Babich0405356,Fergusson08123413,Komatsu09024759,Planck0604069,Komatsu10014538}.
Direct calculation of these correlation functions, however, can be rather involved
\cite{Maldacena0210603,Acquaviva0209156,Seery0503692,Chen0605045,Huang0610235,Arroja08021167,Seery0610210,Arroja09053641,Chen09053494}, as the organization imposed by the slow-roll expansion does not
necessarily
ensure
that the expressions
remain tractable
at intermediate steps.
As such, the underlying
structure behind the final result is obscured.
It would certainly be welcoming to have alternative ways to derive these non-Gaussian correlation functions which emphasize strongly
the symmetries of inflation.
In this regard,
a
promising
guiding principle is
the (approximate) de Sitter symmetry of
 inflation, in which the Hubble
 scale
  $H$ varies slightly during the inflationary evolution.
Since the isometry group of de Sitter spacetime is the 3-dimensional Euclidean conformal group at asymptotic late times, the late time observables, i.e. the correlation functions of primordial fluctuations, are constrained by the (approximate) 3-dimensional conformal symmetry.

The purpose of this paper is to make the above observation precise by
investigating
the implications of the constraints imposed by conformal symmetry.
We have chosen to do so in a language
familiar in the AdS/CFT context,
though
we emphasize that this is merely
terminology that helps organize our study.
We do not make any
reference to
a hypothetical dS/CFT correspondence beyond constraints that can be
derived from symmetry considerations alone.
This holographic
  approach was pioneered in
  \cite{Larsen0202127,vanderSchaar0307271,Larsen0307026} for the
  two-point function.
Here, we shall study the inflationary consistency condition between
the squeezed limit of the three-point function and the two-point
function in this formalism.
Our investigation can be seen as a complementary view to related work in the literature on soft limits using either Ward identities \cite{Assassi12044207} or OPE techniques \cite{Kehagias12051523,Kehagias12101918,Assassi12107792}, on the symmetries of the correlation function of tensor perturbations \cite{Maldacena11042846,Mata:2012bx} or that of
a spectator field \cite{Antoniadis11034164,Creminelli11080874}, on
cosmologies described by broken conformal symmetry
\cite{Hinterbichler11061428,Hinterbichler12026056,Creminelli12034595,Hinterbichler12036351},
on holographically inspired bulk calculations of the bispectrum
\cite{Seery0604209}, and on holographic, strongly coupled
early
universe models
\cite{McFadden10110452,McFadden11043894,Bzowski11121967,Coriano12100136}.\footnote{Note
  that in these last four articles
  \cite{McFadden10110452,McFadden11043894,Bzowski11121967,Coriano12100136}
  {specific} models are considered (all equilateral-shape
  dominated),
whereas the other references and our work here focus
  only on the symmetry constraints within the generic slow-roll context.}

We find that the consistency condition alone already teaches us two valuable lessons:
\begin{enumerate}
\item The slow-roll expansion is not equal to conformal perturbation theory.
\item Ward identities need to be supplemented by evolution equations.
\end{enumerate}
Moreover, from a technical perspective the consistency condition
requires, in the language of conformal field theory, Ward identities
with contributions {beyond linear order:} the inflationary
bispectrum appears as a combination of two- and three-point conformal
correlation functions.
These lessons will be essential
 to analyze the symmetry structure of the full three-point correlation
 function, which we defer to a future work. Appearing shortly after
 this work, \cite{Bzowski:2012ih} showed that in the special case of the slow-roll parameters
 $\epsilon \ll \eta$, one can integrate the conformal perturbation
 series to obtain the full three-point function for this restricted
 class of slow-roll models.

\section{Conformal perturbation theory and the slow-roll expansion}
\begin{figure}[t]
 \centering
 \includegraphics[width=0.60\textwidth]{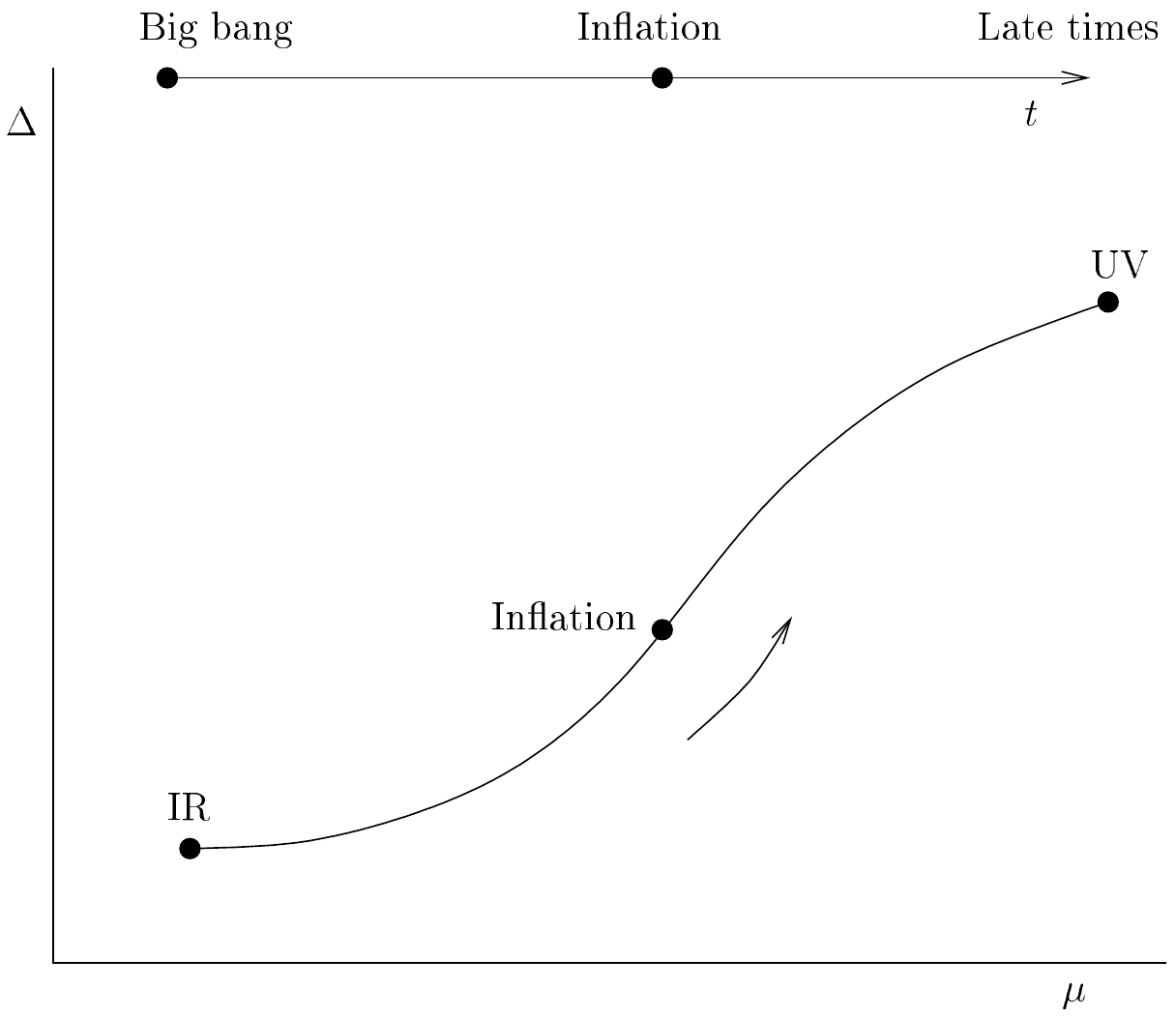}
 \caption{The cosmic evolution can be seen as a reversed renormalization group flow, from the IR fixed point of the dual theory to the UV fixed point of the dual theory.
Inflation occurs at a certain intermediate stage during the renormalization group flow.
As is usual for inflation, a pivot point along the flow is chosen around which the slow-roll expansion can be studied.
We observe the effects of inflation at late times, corresponding to the UV fixed point of the renormalization group flow. \label{fig:RGflow}}
\end{figure}

We briefly review the holographic connection between
conformal field theory and inflationary perturbations
\cite{Maldacena0210603,Larsen0202127,vanderSchaar0307271,Larsen0307026}
(see \cite{McFadden10110452,McFadden11043894} for an extensive
discussion).
The essence of holography is that all calculations of late time
correlation functions of the inflaton can be expressed in terms of a
3-dimensional Euclidean CFT. {The space on which this CFT lives} can be thought of as the asymptotic future of de Sitter space.
Quantitatively,
one identifies the wavefunction of the de Sitter universe {with
appropriate boundary conditions at future infinity}
with
the partition function of the field theory,
\begin{equation}
\label{eq:1}
\Psi_{dS}=Z_{CFT}.
\end{equation}
The asymptotic value $\phi_0(\bvec{x})$ of the inflaton scalar field
$\phi(t,\bvec{x})$, for $t\to\infty$, acts {in the dual CFT} as the coupling $u=\phi_0$ to an operator $\O$.
As a consequence of this coupling, the
conformal field theory $S_{CFT}$, which describes the asymptotic symmetry of {pure} de Sitter spacetime, is perturbed
\begin{equation}\label{eq:BidS:perturbedCFTactiontext}
S_u=S_{CFT}+\int\d^3\bvec{x}\,u\O.
\end{equation}
When the operator is non-marginal, $\Delta=3+\lambda\neq 3$, it will induce a renormalization group flow.
{Compared to} the cosmic evolution this renormalization group flow is reversed:
one
identifies
primordial stages of the cosmic evolution with the IR fixed point of the field theory and late time behavior with the UV fixed point \cite{Strominger0106113,Strominger0110087}.
While one can consider the asymptotic behavior of inflation from the
point of view of the field theory IR fixed point \cite{Larsen0202127},
from the UV fixed point \cite{vanderSchaar0307271} or from the bulk
gravitational IR point of view \cite{Larsen0307026},
it is important to realize that inflation itself actually is an epoch \emph{along} the renormalization group flow, as depicted in figure \ref{fig:RGflow}.

In slow-roll inflation, the weak distortion but still close
  proximity to
de Sitter space due
to the varying inflaton scalar field, is expressed by the requirement that the slow-roll parameters $\epsilon,\eta$ are much smaller than unity but nonzero, $\epsilon,\eta\neq 0$.
From the field theory point of view, it will perturb away
  from the conformal fixed point
$S_{CFT}$ to which the pure de Sitter phase corresponds,
due to the non-marginal nature of $\O$,
i.e. $\lambda=\Delta-3=\pder[\beta]{u}\neq 0$ to first order or, more precisely, its $\beta$ function is non-vanishing, $\beta\neq 0$.
The connection between both sets of perturbation parameters is given via
\begin{equation}\label{eq:BidS:bulkbeta}
\beta=\frac{\partial u}{\partial\mathrm{log}\,\mu}=\frac{\partial\phi}{\partial\mathrm{log}\,a}=\frac{\dot{\phi}}{H}=-2\frac{H'}{ H},
\end{equation}
where the Friedmann equations are used in the last identity and where $\mu=H a$ is the energy scale of the field theory \cite{Larsen0202127,vanderSchaar0307271}.
The slow-roll parameters $\epsilon,\eta$ can then be fully expressed in terms of the CFT-data $\beta,\lambda$,
\begin{align}\label{eq:BidS:slowrollCFT}
\beta^2&=2\epsilon,
&\lambda&=\epsilon-\eta,
&\frac{\partial^2\beta}{\partial u^2}&=\frac{1}{\sqrt{2\epsilon}}\left(2\xi^2-3\epsilon\eta+2\epsilon^2\right).
\end{align}
We have included the third order slow-roll parameter $\xi^2=2\frac{H'H'''}{H^2}$, as it is related to the OPE coefficient $C$ that features prominently in the three-point function of any CFT.

The relation between scaling parameters of the field theory and the slow-roll parameters of the inflationary theory suggests that for the study of slow-roll inflation, for which $\epsilon,\eta\ll 1$, we can consider a nearly marginal deformation of the CFT fixed point, $\beta,\lambda\approx 0$.
It is tempting to identify this with the method of conformal perturbation theory.
This is a known technique, where one can treat general perturbations
in a CFT {abstractly} using only the underlying conformal symmetry \cite{Zamolodchikov87,Cardy:1989da,Freedman0510126}.
Conformal perturbation theory requires a small deviation from
marginality $\lambda\ll 1$ {and} a small coupling $u\ll 1$, plus it
always starts from a fixed point.
Using conformal perturbation theory, phrased with respect to the unperturbed conformal weight $\Delta_0=\Delta-\gamma$ where $\gamma$
is the anomalous dimension\footnote{Recall
  that the value of the dimension of an operator changes along the
  RG-flow. Because conformal perturbation theory naturally starts from
  a fixed point, but inflation is a point along the RG-flow, this
  distinction between the fixed point value $\Delta_0$ and the real
  value $\Delta(u)$ of the scaling dimension is important as we
  emphasized in figure \ref{fig:RGflow}.}, the $\beta$ function can be given as a perturbative expression in $u$ \cite{Zamolodchikov87,Cardy:1989da,Freedman0510126},
\begin{equation}\label{eq:string:betaConfPertTheory}
\beta(u)=(\Delta_{0}-3)u+2\pi Cu^2+\ldots=\lambda u+\ldots.
\end{equation}
However, since slow-roll expansion
is an expansion around an intermediate point of the RG flow, the identification between conformal perturbation theory and slow-roll inflation is not exact.
E.g. one issue that surfaces  when comparing \eqref{eq:BidS:slowrollCFT} and \eqref{eq:string:betaConfPertTheory}
 is that as $\beta^2=O(\epsilon)$ and $\lambda=O(\epsilon)$, these
  imply that $u$ itself is of order $O(\epsilon^{-1/2})$.
If any, the slow-roll expansion seems to correspond to the \emph{large} $u$-regime.
The
relation between the slow-roll expansion and conformal perturbation theory thus
 appears to be
rather
subtle.

\section{Ward identities}
Using the constraints of {late time asymptotic} conformal symmetry, previous studies have been able to completely understand a spectator field $\phi$ in an inflationary background \cite{Antoniadis11034164,Creminelli11080874}.
Here we want to study the adiabatic mode of curvature density perturbations $\zeta$.
This is holographically dual to the trace $\Theta$ of the
stress-energy tensor of the
perturbed CFT, see e.g.~\cite{McFadden10110452,McFadden11043894}.
The precise relation between $\zeta$- and $\Theta$-correlation
functions follows from a standard expansion of the wavefunction/partition function \eqref{eq:1} \cite{Maldacena0210603,Maldacena11042846},
\begin{align}\label{eq:BidS:holographicnpt}
\langle\zeta_{\bvec{k}}\zeta_{-\bvec{k}}\rangle'&\propto\frac{-1}{\mathrm{Re}\langle \Theta_{\bvec{k}} \Theta_{-\bvec{k}}\rangle'},
&\langle\zeta_{\bvec{k}_1}\zeta_{\bvec{k}_2}\zeta_{\bvec{k}_3}\rangle&\propto\frac{-\mathrm{Re}\langle \Theta_{\bvec{k}_1} \Theta_{\bvec{k_2}} \Theta_{\bvec{k_3}}\rangle}{\prod_{j=1}^3\mathrm{Re}\langle \Theta_{\bvec{k}_j} \Theta_{-\bvec{k}_j}\rangle'}.
\end{align}
Here a prime $'$ indicates that we consider the part of the correlation function that multiplies the momentum-conserving delta function.
The important difference between a spectator field and the adiabatic
mode $\zeta$ is that the trace of the stress-energy tensor is not a
standard primary operator of the CFT.
As a result, we cannot calculate \eqref{eq:BidS:holographicnpt}
directly by exploiting the restricted form of correlation functions of
CFT primary operators.
Nevertheless, because $\Theta$ is the generator of the Weyl symmetries, we can resort to Ward identities between $\Theta$ and $\O$ to obtain the correlation functions of $\Theta$.

In the gravity calculation the curvature perturbations are defined in a gauge independent way and one has the freedom to choose a specific gauge in which the calculation is done.
The gauge invariance of the gravity theory is translated to the fact that in a scale dependent field theory one can either change the dimensionful coupling $u$ to the operator $\O$ or one can change the metric, relating $\zeta$ to $\Theta$ accordingly \cite{Maldacena0210603}.
This gauge relation is reflected in the CFT, since both operators are
related \emph{to leading order}\footnote{This operator identity is
  only valid up to local terms in correlation functions.} via
\cite{Bianchi0310129,Muck10062987}
\begin{equation}\label{eq:BidS:ThetabetaOeasy}
\Theta=-\beta\O,
\end{equation}
where the standard renormalization $\beta$ function serves as the constant of proportionality.

In general, gauge symmetries of a theory
correspond to
constraints.
In the case of the gravity theory, these are the hamiltonian and momentum/reparameterization constraints of the ADM formalism  \cite{Maldacena11042846}.
In a field theory, the symmetries impose constraints on the correlation function through Ward identities.
In our case, we need to find the relations between the two- and three-point functions of $\Theta$ and $\O$.
As the trace of the stress-energy tensor is the generator of Weyl transformations, we consider the Ward identities of (multiple) trace insertions.
Initially the calculation follows directly from a textbook field theory calculation \cite{DiFrancesco:1997nk}, but once multiple trace insertions have to be taken into account, more care is required, as subleading, local contributions other than (powers of) \eqref{eq:BidS:ThetabetaOeasy} contribute (see also \cite{McFadden10110452,McFadden11043894}),
\begin{subequations}\label{eq:BidS:Wardtext}
\begin{align}
\cor{\Theta_u(\bvec{x})}&=-u\lambda\cor{\O(\bvec{x})},\label{eq:BidS:Ward1text}\\
\cor{\Theta_u(\bvec{x})\Theta_u(\bvec{y})}
&=u^2\lambda^2\cor{\O(\bvec{x})\O(\bvec{y})}-u\lambda^2\delta (\bvec{x}-\bvec{y})\cor{\O(\bvec{x})},\label{eq:BidS:Ward2text}\\
\cor{\Theta_u(\bvec{x})\Theta_u(\bvec{y})\Theta_u(\bvec{z})}
&=-u^3\lambda^3\cor{\O(\bvec{x})\O(\bvec{y})\O(\bvec{z})}\label{eq:BidS:Ward3text}\\
&\phantom{=}+u^2\lambda^3\delta (\bvec{y}-\bvec{z})\cor{\O(\bvec{x})\O(\bvec{y})}+\bvec{x}\leftrightarrow \bvec{y}+\bvec{y}\leftrightarrow \bvec{z}+\ldots.\notag
\end{align}
\end{subequations}
The $\ldots$ contain highly local contributions that are negligible
for our purposes and the subscript $\langle\ \rangle_u$ on the correlation
  function means that they are evaluated in the perturbed CFT \eqref{eq:BidS:perturbedCFTactiontext}.
We refer to the appendix for details regarding the derivation of \eqref{eq:BidS:Wardtext}.

Note, the Ward identities \eqref{eq:BidS:Wardtext} are valid \emph{throughout} the renormalization group flow, i.e. for each value of $u$.
Although the familiar $\beta(u)$-dependence from \eqref{eq:BidS:ThetabetaOeasy} seems missing, \eqref{eq:BidS:Wardtext} does contain the $u$-dependent conformal weight $\Delta(u)=3+\lambda$, which carries similar information \cite{Bianchi0310129,Muck10062987}.
In fact, in the conformal perturbation theory limit $u\to 0$, one can expand the one-point correlation function $\cor{\O}$ with respect to the unperturbed theory $\cor[0]{\O}$, to find \eqref{eq:BidS:ThetabetaOeasy} with $\beta=(\Delta_0-3)u+\ldots$ given by \eqref{eq:string:betaConfPertTheory}.
Next note that even though $u$ is of order $O(\epsilon^{-1/2})$, the combination $Cu$ appearing at subleading order in \eqref{eq:string:betaConfPertTheory}, is of order $O(\epsilon)$ and appears with increasing power, $Cu$, $(Cu)^2$, etc., for higher orders in $u$.
Hence, although we cannot be certain of the validity of conformal perturbation theory itself, the expansion of perturbed correlation functions in terms of unperturbed correlation functions is very much similar to the slow-roll expansion.

\section{Consistency condition}
From \eqref{eq:BidS:Ward2text} and the corresponding expression in conformal perturbation theory,
\begin{equation}
\cor{\Theta_u\Theta_u}=\beta^2\cor[0]{\O\O},
\end{equation}
the usual power spectrum of adiabatic curvature fluctuations can be
obtained by solving the Callan-Symanzik equation for
  $\cor{\Theta_u\Theta_u}$ \cite{vanderSchaar0307271}.
As is clear from the analysis in \cite{vanderSchaar0307271}, the use of conformal perturbation theory does not immediately give an intuitively correct result.
Conformal perturbation theory only allows us to start from a fixed point of the renormalization group flow, whereas the curvature power spectrum is expressed with respect to the classical inflationary evolution, driven by an almost ---but not exactly--- constant Hubble parameter $H(u)$.
Although in the case of the power spectrum this can be remedied by
explicitly including an appropriate expression for $H(u)$, it does
show that the difference between conformal perturbation
theory and the slow-roll expansion is not just mathematically subtle.
Its intricacy has physical consequences as well.

The next obvious object to consider is the three-point correlation function.
The derivation of the full bispectrum using holography will be addressed in a future work.
Here,
we focus on obtaining the inflationary consistency relation holographically, as an important first step {in} this
direction.
The consistency condition in the squeezed limit is given by \cite{Maldacena0210603,Creminelli0407059,Cheung07090295},
\begin{equation}\label{eq:BidS:squeezedlimit}
\lim_{k_3\to 0} \langle\zeta_{\bvec{k}_1}\zeta_{\bvec{k}_2}\zeta_{\bvec{k}_3}\rangle=-\delta (\bvec{k}_1+\bvec{k}_2+\bvec{k}_3)
\langle\zeta_{\bvec{k}_1}\zeta_{-\bvec{k}_1}\rangle' \langle\zeta_{\bvec{k}_3}\zeta_{-\bvec{k}_3}\rangle'
\frac{\d\log k_1^3\langle\zeta_{\bvec{k}_1}\zeta_{-\bvec{k}_1}\rangle' }{\d\log k_1},
\end{equation}
which we wish to reproduce from \eqref{eq:BidS:Ward3text}.
Our analysis shows that the Ward identity needs to be supplemented by a dynamical equation dictating the inflationary evolution.

To do so
note that the Fourier transform of the second term(s) on the
  RHS in \eqref{eq:BidS:Ward3text} is given by
\begin{align}\label{eq:BidS:OOcontribution}
\delta (\bvec{x}_2-\bvec{x}_3)\cor[0]{\O(\bvec{x}_1)\O(\bvec{x}_2)}&
\xrightarrow{\mathrm{F.T.}}
\delta (\bvec{k}_1+\bvec{k}_2+\bvec{k}_3)\cor[0]{\O_{\bvec{k}_1}\O_{-\bvec{k}_1}}'.
\end{align}
The overall momentum-conserving delta function can therefore be divided out on both sides of equation \eqref{eq:BidS:Ward3text}.
In the limit $k_3\to 0$, the Fourier transform of the first line
on the RHS of \eqref{eq:BidS:Ward3text} gives
\begin{align}
\lim_{k_3\to 0}\cor[0]{\O_{\bvec{k}_1}\O_{\bvec{k}_2}\O_{\bvec{k}_3}}'
&=\int\d^3\bvec{x}\cor[0]{\O_{\bvec{k}_1}\O_{-\bvec{k}_1}\O(\bvec{x})}'
=-\frac{\partial}{\partial u}\cor{\O_{\bvec{k}_1}\O_{-\bvec{k}_1}}'+O(u).
\end{align}
Now we use the Callan-Symanzik equation applied to the two-point function \cite{Callan:1970yg,Symanzik:1970rt,Symanzik:1971vw},
\begin{equation}
\left(k\frac{\partial}{\partial k}-\beta\frac{\partial}{\partial u}-2\lambda -3\right)\cor{\O_{\bvec{k}}\O_{-\bvec{k}}}'=0,
\end{equation}
to rewrite this expression as
\begin{align}
\lim_{k_3\to 0}\cor[0]{\O_{\bvec{k}_1}\O_{\bvec{k}_2}\O_{\bvec{k}_3}}'
&=\frac{-1}{\beta}\left(k_1\frac{\partial}{\partial k_1}-3\right)\cor[0]{\O_{\bvec{k}_1}\O_{-\bvec{k}_1}}'+\frac{2}{\beta}\lambda\cor[0]{\O_{\bvec{k}_1}\O_{-\bvec{k}_1}}'+O(u).
\end{align}
In the conformal perturbation theory limit, where $\beta=\lambda u$ to lowest order, the squeezed limit of the Ward identity yields,
\begin{align}
\lim_{k_3\to 0}\cor{\Theta_u(\bvec{k}_1)\Theta_u(\bvec{k}_2)\Theta_u(\bvec{k}_3)}'
&=\beta^2\left(k_1\frac{\partial}{\partial k_1}-3\right)\cor[0]{\O_{\bvec{k}_1}\O_{-\bvec{k}_1}}'-2\beta^2\lambda\cor[0]{\O_{\bvec{k}_1}\O_{-\bvec{k}_1}}'\notag\\
&\phantom{=}+\beta^2\lambda\left(2\cor[0]{\O_{\bvec{k}_1}\O_{-\bvec{k}_1}}'+0\right),
\end{align}
where we have used that $\lim_{k_3\to 0}\cor[0]{\O_{\bvec{k}_3}\O_{-\bvec{k}_3}}'=0$.
The cancelation of the second line in \eqref{eq:BidS:Ward3text}
against the anomalous contribution from the Callan-Symanzik equation
ensures that \eqref{eq:BidS:holographicnpt} satisfies the squeezed
limit consistency condition \eqref{eq:BidS:squeezedlimit},
\begin{align}
\lim_{k_3\to 0}\cor[]{\zeta_{\bvec{k}_1}\zeta_{\bvec{k}_2}\zeta_{\bvec{k}_3}}'
&\propto\frac{\left(k_1\frac{\partial}{\partial k_1}-3\right)\cor[]{\O_{\bvec{k}_1}\O_{-\bvec{k}_1}}'}{\beta^4\cor[]{\O_{\bvec{k}_1}\O_{-\bvec{k}_1}}'\cor[]{\O_{\bvec{k}_2}\O_{-\bvec{k}_2}}'\cor[]{\O_{\bvec{k}_3}\O_{-\bvec{k}_3}}'}\notag\\
&=\cor[]{\zeta_{\bvec{k}_1}\zeta_{-\bvec{k}_1}}'^2\cor[]{\zeta_{\bvec{k}_3}\zeta_{-\bvec{k}_3}}'
\left[-\frac{1}{\cor[]{\zeta_{\bvec{k}_1}\zeta_{-\bvec{k}_1}}'} \left(k_1\frac{\partial}{\partial k_1}\log\cor[]{\zeta_{\bvec{k}_1}\zeta_{-\bvec{k}_1}}'+3\right)\right].
\end{align}

A
separate, independent derivation of the consistency condition \eqref{eq:BidS:squeezedlimit} using similar ingredients
has been given in \cite{Assassi12044207}, in which the (broken) conformal symmetry is
described
using a Ward identity.
This Ward identity is equivalent to the Callan-Symanzik equation in
  our formalism,\footnote{Recall that the Callan-Symanzik equation is
    essentially the Ward identity for $\cor[u]{\Theta{\cal
        O}\ldots}=\mu\frac{\partial}{\partial \mu} \cor[u]{{\cal
        O}\ldots}$
    \cite{Callan:1970yg}.} whereas our Ward identity relating
  $\Theta$ and $\O$ has no equivalent in the description of
  \cite{Assassi12044207}, which work directly with the gauge-invariant
  curvature perturbation $\zeta$.
Since Ward identities naturally relate an $n+1$-point correlation function with the variation of an $n$-point function, the observation in \cite{Hinterbichler12036351,Assassi12044207} is that the consistency condition essentially \emph{is} a Ward identity, applied to a particular conserved current.
The current under consideration corresponds to a combination of a shift and dilational transformation, perturbing the system much in the same way as the Callan-Symanzik equation in our formalism.
It would be very interesting to
further investigate the connection between \cite{Assassi12044207} and our work.

\section{Outlook}
Using conformal techniques, our derivation of the consistency
condition in the squeezed limit has taught us valuable lessons about
the structure of the three-point function.
\begin{enumerate}
\item
We have seen that one has to be very careful in drawing a parallel between
slow-roll expansion
and conformal perturbation theory
even though
naively
they are
related.
\item
We have also seen that in order to apply conformal
  techniques and Ward identities to
inflationary perturbations, an essential ingredient is the use of the evolution equation to
take into account
 the inflationary evolution.
\end{enumerate}
These two points are also clear from the article \cite{Bzowski:2012ih}
which appeared shortly after this work. 1. As we do, they emphasize that
the main difference between conformal perturbation theory and slow-roll is
that conformal perturbation theory starts from a fixed point whereas
slow-roll is defined at an arbitrary point in the evolution. There is
therefore a special slow-roll case
where direct conformal perturbation theory should give the
answer immediately. This is a slow-roll inflation scenario {\em very} close to a
fixed point. 2. The authors of \cite{Bzowski:2012ih} use this {twice} for inflationary models interpolating between two very close
fixed points, such that the regimes of conformal perturbation theory
from each overlap. In
terms of cosmological parameters, this corresponds to the limit
where $\epsilon \ll \eta$.
This allows them to {solve} for the evolution. In this special
case they are then able to obtain
the full three-point function, which clearly obeys the consistency
condition \cite{Bzowski:2012ih}. In contrast, we use the abstract evolution
equation to show that a holographic derivation of the consistency
relation is always true for any slow-roll model.

\begin{figure}[t]
 \centering
 \includegraphics[width=0.60\textwidth]{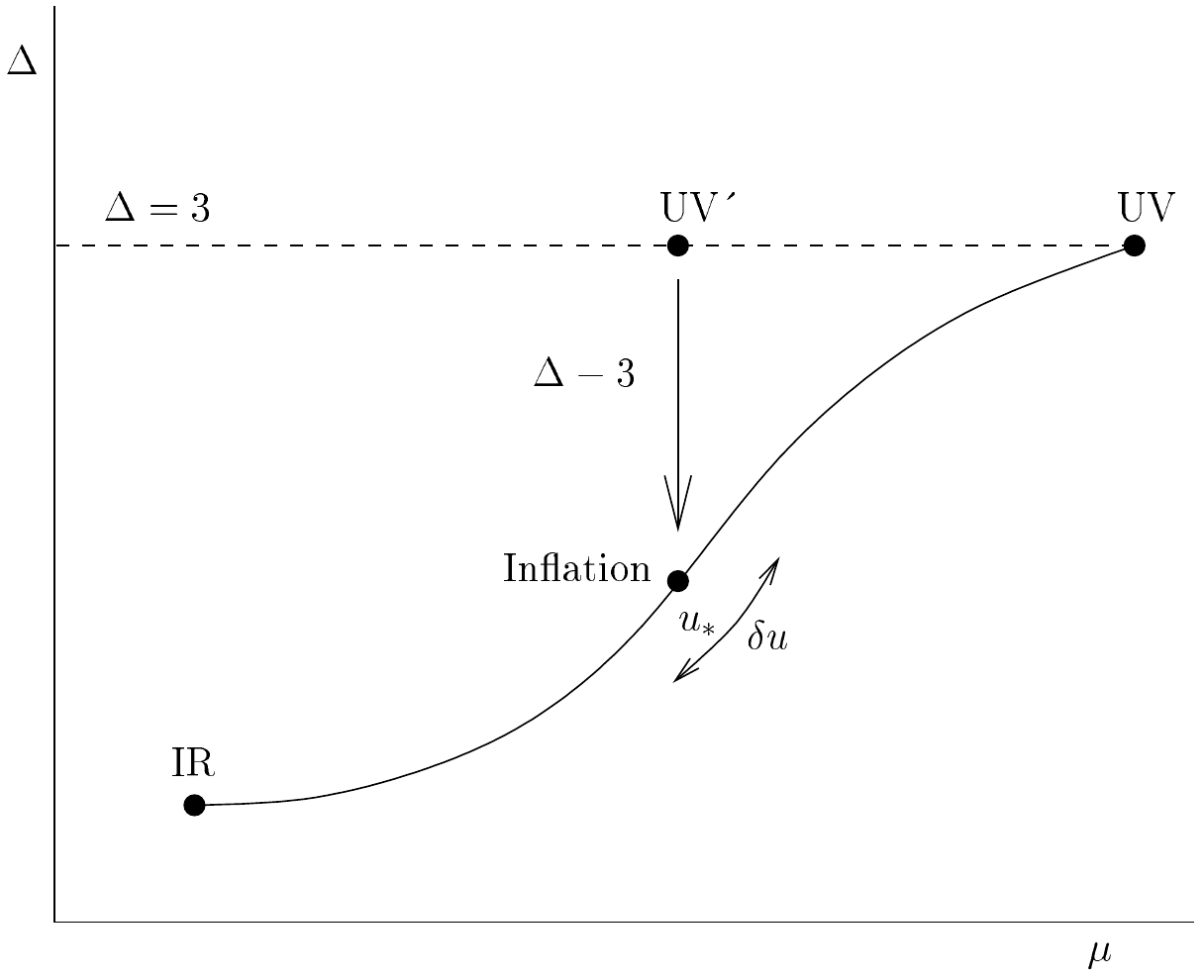}
 \caption{The inflationary phase at an intermediate point in the renormalization group flow may be approximated by a conformal fixed point.
 The dashed line indicates a marginal renormalization group flow from one UV theory to another, for an operator with exactly marginal dimension $\Delta=3$.
 The validity of the slow-roll approximation suggests that expressions in the quasi-fixed point can be approximated by a $(\Delta-3)$-expansion.
\label{fig:quasidS}}
\end{figure}

These insights provide useful hints on how to proceed
to derive the full bispectrum for the general slow-roll case
as well as for higher $n$-point functions.
The squeezed limit is a relative statement, $k_3\ll k_1,k_2$, focussing on the high frequency limit $k_1\to\infty$.
    Hence, in the full three-point function we expect UV effects, i.e. counterterms arising from UV divergences, to be a dominant contribution as well.
Secondly, the explicit $\frac{1}{\beta}$ necessary for the cancelation against the two-point correlation function contribution of \eqref{eq:BidS:Ward3text} indicates that non-analytic behavior in $\beta$ plays an essential role.
Via $\beta=\lambda u$, this can be investigated through an expansion
in marginality $\lambda$.
The inflationary evolution, dual to the renormalization group flow at
an intermediate stage, should be approximated by a quasi-conformal
fixed point $\lambda=\epsilon-\eta\approx 0$ because of the slow-roll
approximation, cf.~figure \ref{fig:quasidS}.
By analyzing the Laurent series of $\cor[0]{\O\O\O}$ in $\lambda$, one should be able to uncover the dominant contributions.

\acknowledgments
We would like to thank Daniel Baumann, Paolo Creminelli, Guido D'Amico, Aki Hashimoto, Kurt Hinterbichler, Justin Khoury, Yoske Sumitomo and Peng Zhou
for interesting discussions and interactions, and especially Paul McFadden and
Kostas Skenderis in comparing our respective results.
TvdA is grafeful to the Department of Physics of the University of Wisconsin for their support and hospitality in the initial phase of this research.
We would like to thank the Lorentz Center for their support and the participants and organizers of the workshop ``Effective field theory in inflation'' for stimulating discussions during the completing stages.
The research of KS and TvdA was supported
in part by a VIDI Innovative Research Incentive Grant from the Netherlands Organisation for Scientific Research (NWO) and by the Dutch Foundation for
Fundamental Research on Matter (FOM).
The work of GS was supported in part by DOE grant DE-FG-02-95ER40896. GS would also like to thank the University of Amsterdam for hospitality during part of this work, as he was visiting the Institute for Theoretical Physics as the Johannes Diderik van der Waals Chair.

\appendix
\section{Ward identities of multiple trace insertions\label{sec:BidS:Wardapp}}
The Ward identity of any symmetry generator can be derived from considering infinitesimal transformations of the correlation function
\begin{equation}\label{eq:BidS:pathintegralcorrelation}
 \cor{X}=\frac{1}{Z}\int\mathcal{D}\O\,Xe^{-S_u[\O]},
\end{equation}
of a product of operators $X=\O(\bvec{x}_1)\ldots\O(\bvec{x}_n)$.
We specifically evaluate the expectation value with respect to a \emph{perturbed} conformal field theory,
\begin{equation}\label{eq:BidS:perturbedCFTaction}
S_u[\O]=S_{CFT}[\O]+\int\d^3 \bvec{x}\,u\O(\bvec{x}).
\end{equation}
Under the transformation $\O\to\O'=\O(\bvec{x})-i\omega_a G_a(\bvec{x})\O(\bvec{x})$, the action transforms as
\begin{align*}
S_u[\O']&=S_u[\O]-\int\d^3 \bvec{x}\,\omega_a(\bvec{x})\partial_\mu j_a^\mu(\bvec{x})-i\int\d^3 \bvec{x}\,\omega_a(\bvec{x})G_a(\bvec{x})u\O(\bvec{x})\notag\\
&\phantom{=S_u[\O]}-\frac12\int\d^3 \bvec{x}\, \omega_a(\bvec{x})G_a(\bvec{x})\omega_b(\bvec{x})G_b(\bvec{x})u\O(\bvec{x})+\ldots,
\end{align*}
to second order in $\omega_a$, where $j_a^\mu$ is the Noether current of the conformal field theory $S_{CFT}[\O]$ at the fixed point.
The last term is a contact term, which we have included because it is of second order in $\omega_a$.
It stems from the transformation $\O\to\O'$ in the perturbation-part of the action \eqref{eq:BidS:perturbedCFTaction}.
In principle, the unperturbed conformal action $S_{CFT}[\O]$ also obtains a contribution at second order as a result of the transformation to second order.
However, this contribution is difficult to retrieve from first principles, as the Noether current of the transformation is only defined infinitesimally.
It is therefore left implicit in the $\ldots$, while its effect on the Ward identity will later be inferred by different means.
Hence, for the moment we find
\begin{align*}
e^{-S_u[\O']}=e^{-S_u[\O]}&\Big(1-\int\d^3 \bvec{x}\,\omega_a(\bvec{x}) \delta L_a(\bvec{x})+\frac12\int\d^3 \bvec{x}\d \bvec{y}\,\omega_a(\bvec{x})\omega_b(\bvec{y})\delta L_a(\bvec{x})\delta L_b(\bvec{y})\Big)\notag\\
\times\, &\Big(1+\frac12\int\d^3 \bvec{x}\, \omega_a(\bvec{x})G_a(\bvec{x})\omega_b(\bvec{x})G_b(\bvec{x})u\O(\bvec{x})\Big),
\end{align*}
up to second order,
where $\delta L_a(\bvec{x})$ is shorthand notation for
\begin{equation*}
\delta L_a(\bvec{x})=-\partial_\mu j^\mu_a(\bvec{x})-iuG_a(\bvec{x})\O(\bvec{x}).
\end{equation*}
Infinitesimally transforming $X$ gives
\begin{align*}
X'=e^{-i\omega_a(\bvec{x})G_a(\bvec{x})}X=&X-i\sum_k\omega_{a,k}G_{a,k}X
-\frac12\sum_{k,l}\omega_{a,k}G_{a,k}\omega_{b,l}G_{b,l}X,
\end{align*}
where $\omega_{a,k}=\omega_a(\bvec{x}_k)$ and $G_{a,k}=G_a(\bvec{x}_k)$ acts on the $k$'th $\O(\bvec{x}_k)$ inside $X$.

Assuming the measure is invariant, $\mathcal{D} \O'=\mathcal{D} \O$, comparison of the transformed expression for $\cor{X}$ and \eqref{eq:BidS:pathintegralcorrelation} gives,
\begin{align}
0&=\int\d^3 \bvec{x}\, \omega_a(\bvec{x})\cor{\delta L_a(\bvec{x})X}+i\sum_k\omega_{a,k}G_{a,k}\cor{X},\label{eq:BidS:WardL}
\end{align}
to first order in $\omega$.
Similarly, to second order it gives
\begin{align}
0&
=\int\d^3 \bvec{x}\d^3 \bvec{y}\,\omega_a(\bvec{x})\omega_b(\bvec{y})\cor{\delta L_a(\bvec{x})\delta L_b(\bvec{y})X}
+2i\sum_k\omega_{a,k}G_{a,k}\int\d^3 \bvec{x}\,\omega_b(\bvec{x}) \cor{\delta L_b(\bvec{x})X}\notag\\
&\phantom{=}-\sum_{k,l}\omega_{a,k}G_{a,k}\omega_{b,l}G_{b,l}\cor{X}
+\int\d^3\bvec{x}\,\omega_a(\bvec{x})G_a(\bvec{x})\omega_b(\bvec{x})G_b(\bvec{x})u\cor{\O(\bvec{x})X}\notag\\
&=\int\d^3 \bvec{x}\d^3 \bvec{y}\,\omega_a(\bvec{x})\omega_b(\bvec{y})\cor{\delta L_a(\bvec{x})\delta L_b(\bvec{y})X}
+\sum_{k,l}\omega_{a,k}G_{a,k}\omega_{b,l}G_{b,l}\cor{X}\notag\\
&\phantom{=}+\int\d^3\bvec{x}\,\omega_a(\bvec{x})G_a(\bvec{x})\omega_b(\bvec{x})G_b(\bvec{x})u\cor{\O(\bvec{x})X}.\label{eq:BidS:WardLL}
\end{align}
We consider the special choice for $\omega_a=\omega(\bvec{x})(1_D,-x^\nu1_T)$ to derive the trace insertion formulae, where $1_D$ and $1_T$ mean that we consider the dilational and translational transformations.
The combined effect of this familiar combination
\cite{CallanColemanJackiw70,Assassi12044207,Underwood10094200} yields
\begin{align*}
\omega_a(\bvec{x})G_a(\bvec{x})&=\omega(\bvec{x})\left(-i(x^\mu\partial_\mu+\Delta)-x^\nu\left(-i\partial_\nu\right)\right)=-i\omega(\bvec{x})\Delta,\notag\\
\omega_a(\bvec{x})\delta L_a(\bvec{x})&=\omega(\bvec{x})\left(-\Theta_0(\bvec{x})-u\Delta\O(\bvec{x})\right)=\omega(\bvec{x})\left(-\Theta_u(\bvec{x})-u(\Delta-3)\O(\bvec{x})\right).
\end{align*}
In the last line we rewrite the answer in terms of the stress-energy tensor of the \emph{perturbed} theory,
\begin{equation}
\Theta_u=\left.\frac{-2}{\sqrt{h}}h^{\mu\nu}\frac{\delta S}{\delta h^{\mu\nu}}\right|_{h_{\mu\nu}=\delta_{\mu\nu}}=\Theta_0+\frac{-2}{\sqrt{h}}h^{\mu\nu}\left.\frac{-1}{2}\sqrt{h}h_{\mu\nu}\right|_{h_{\mu\nu}=\delta_{\mu\nu}}u\O=\Theta_0+3u\O.
\end{equation}
The Ward identities can then be written as
\begin{align}\label{eq:BidS:Ward1}
\cor{\Theta_u(\bvec{x})X}&=-u(\Delta-3)\cor{\O(\bvec{x})X}+\sum_k\delta (\bvec{x}-\bvec{x}_k)\Delta_k\cor{X},
\end{align}
and
\begin{align}\label{eq:BidS:Ward2}
\cor{\Theta_u(\bvec{x})\Theta_u(\bvec{y})X}
&=\sum_{k,l}\delta (\bvec{x}-\bvec{x}_k)\Delta_k\delta (\bvec{y}-\bvec{x}_l)\Delta_l\cor{X}
+u\delta (\bvec{x}-\bvec{y})\Delta^2\cor{\O(\bvec{x})X}\notag\\
&\phantom{=}-u(\Delta-3)\cor{\Theta_u(\bvec{x})\O(\bvec{y})X}-\bvec{x}\leftrightarrow\bvec{y}
-u^2(\Delta-3)^2\cor{\O(\bvec{x})\O(\bvec{y})X}\notag\\
&=\sum_{k,l}\delta (\bvec{x}-\bvec{x}_k)\Delta_k\delta (\bvec{y}-\bvec{x}_l)\Delta_l\cor{X}\notag\\
&\phantom{=}+u\delta (\bvec{x}-\bvec{y})\left(\Delta^2-2(\Delta-3)\Delta\right)\cor{\O(\bvec{x})X}\notag\\
&\phantom{=}-u(\Delta-3)\sum_{k}\delta (\bvec{x}-\bvec{x}_k)\Delta_k\cor{\O(\bvec{y})X}-\bvec{x}\leftrightarrow\bvec{y}\notag\\
&\phantom{=}+u^2(\Delta-3)^2\cor{\O(\bvec{x})\O(\bvec{y})X},
\end{align}
where $\Delta_k$ is the (full) scaling dimension of the $k$'th operator $\O(\bvec{x}_k)$ inside $X$.

At this stage, we have to reflect on the correctness of the expressions by performing a consistency check on the two-point function \eqref{eq:BidS:Ward2}.
When we perturb the conformal field theory with a purely marginal operator, $\Delta=3$, the renormalization group flow remains in a (different) conformal field theory.
The trace $\Theta_u$ of this perturbed theory is still vanishing.
Hence, the two-point correlation function of the perturbed stress-energy tensor should vanish with respect to the perturbed theory.
However, substituting $\Delta= 3$ into our expression,
\begin{equation*}
\cor{\Theta_u(\bvec{x})\Theta_u(\bvec{y})}=u^2(\Delta-3)^2\cor{\O(\bvec{x})\O(\bvec{y})}-u\delta (\bvec{x}-\bvec{y})\left((\Delta-3)^2-3^2\right)\cor{\O(\bvec{x})}, \end{equation*}
does not yield zero.
Clearly in our derivation we must have missed a term of the form $-3^2u\delta (\bvec{x}-\bvec{y})\cor{\O(\bvec{x})}$.
This term is a contact term and
should stem
from the neglected second order transformation of $S_{CFT}[\O]$, which will contain a contribution from the variation of the Noether current.
In \cite{McFadden10110452,McFadden11043894,Bzowski11121967} such a contribution is explicitly included for the consistency of the expressions.
In our case
we can infer the final result based on conceptual
reasoning. We thus employ the expression
\begin{align}\label{eq:BidS:Ward2corrected}
\cor{\Theta_u(\bvec{x})\Theta_u(\bvec{y})X}
&=u^2(\Delta-3)^2\cor{\O(\bvec{x})\O(\bvec{y})X}-u(\Delta-3)^2\delta (\bvec{x}-\bvec{y})\cor{\O(\bvec{x})X}\notag\\
&\phantom{=}-u(\Delta-3)\sum_{k}\delta (\bvec{x}-\bvec{x}_k)\Delta_k\cor{\O(\bvec{y})X}-\bvec{x}\leftrightarrow\bvec{y}\notag\\
&\phantom{=}+\sum_{k,l}\delta (\bvec{x}-\bvec{x}_k)\Delta_k\delta(\bvec{y}-\bvec{x}_l)\Delta_l\cor{X},
\end{align}
for the double Ward identity.

We similarly
derive the correlation function of three traces.
Evaluating the above expressions to third order in $\omega$, one obtains the Ward identity of three current insertions,
\begin{align}
0&=\int\d^3 \bvec{x}\d^3 \bvec{y}\d^3 \bvec{z}\,\omega_a(\bvec{x})\omega_b(\bvec{y})\omega_c(\bvec{z})\cor{\delta L_a(\bvec{x})\delta L_b(\bvec{y})\delta L_c(\bvec{z})X}\notag\\
&-i\sum_{k,l,m}\omega_{a,k}G_{a,k}\omega_{b,l}G_{b,l}\omega_{c,m}G_{c,m}\cor{X}+\ldots,
\end{align}
where the $\ldots$ contain terms of order $u$.
These terms are double contact terms and, as we have seen in the main text, are not relevant for our purposes.
Choosing again $\omega_a=\omega(\bvec{x})(1_D,-x^\nu1_T)$ and using the single and double trace-inserted Ward identities (\ref{eq:BidS:Ward1}), (\ref{eq:BidS:Ward2corrected}), the three-point function of the stress-energy tensor reads
\begin{align}\label{eq:BidS:Ward3}
\cor{\Theta_u(\bvec{x})\Theta_u(\bvec{y})\Theta_u(\bvec{z})X}
&=-u^3(\Delta-3)^3\cor{\O(\bvec{x})\O(\bvec{y})\O(\bvec{z})X}\notag\\
&\phantom{=}-u^2(\Delta-3)^2\cor{\Theta_u(\bvec{x})\O(\bvec{y})\O(\bvec{z})X}-\bvec{x}\leftrightarrow \bvec{y}-\bvec{x}\leftrightarrow \bvec{z}\notag\\
&\phantom{=}-u(\Delta-3)\cor{\Theta_u(\bvec{x})\Theta_u(\bvec{y})\O(\bvec{z})X}-\bvec{z}\leftrightarrow \bvec{x}-\bvec{z}\leftrightarrow \bvec{y}+\ldots\notag\\
&=-u^3(\Delta-3)^3\cor{\O(\bvec{x})\O(\bvec{y})\O(\bvec{z})X}\notag\\
&\phantom{=}+u^2(\Delta-3)^3\delta (\bvec{y}-\bvec{z})\cor{\O(\bvec{x})\O(\bvec{y})X}+\bvec{x}\leftrightarrow \bvec{y}+\bvec{y}\leftrightarrow \bvec{z}+\ldots,
\end{align}
where the $\ldots$ contain highly local contributions.


\begin{thebibliography}{10}

\bibitem{Maldacena0210603}
J.~M. Maldacena, {\it {Non-Gaussian features of primordial fluctuations in
  single field inflationary models}},  {\em JHEP} {\bf 0305} (2003) 013,
  [\href{http://xxx.lanl.gov/abs/astro-ph/0210603}{{\tt astro-ph/0210603}}].

\bibitem{Acquaviva0209156}
V.~Acquaviva, N.~Bartolo, S.~Matarrese, and A.~Riotto, {\it {Second order
  cosmological perturbations from inflation}},  {\em Nucl.Phys.} {\bf B667}
  (2003) 119--148, [\href{http://xxx.lanl.gov/abs/astro-ph/0209156}{{\tt
  astro-ph/0209156}}].

\bibitem{Seery0503692}
D.~Seery and J.~E. Lidsey, {\it {Primordial non-Gaussianities in single field
  inflation}},  {\em JCAP} {\bf 0506} (2005) 003,
  [\href{http://xxx.lanl.gov/abs/astro-ph/0503692}{{\tt astro-ph/0503692}}].

\bibitem{Chen0605045}
X.~Chen, M.-x. Huang, S.~Kachru, and G.~Shiu, {\it {Observational signatures
  and non-Gaussianities of general single field inflation}},  {\em JCAP} {\bf
  0701} (2007) 002, [\href{http://xxx.lanl.gov/abs/hep-th/0605045}{{\tt
  hep-th/0605045}}].

\bibitem{Huang0610235}
X.~Chen, M.-x. Huang, and G.~Shiu, {\it {The Inflationary Trispectrum for
  Models with Large Non-Gaussianities}},  {\em Phys.Rev.} {\bf D74} (2006)
  121301, [\href{http://xxx.lanl.gov/abs/hep-th/0610235}{{\tt
  hep-th/0610235}}].

\bibitem{Arroja08021167}
F.~Arroja and K.~Koyama, {\it {Non-gaussianity from the trispectrum in general
  single field inflation}},  {\em Phys.Rev.} {\bf D77} (2008) 083517,
  [\href{http://xxx.lanl.gov/abs/0802.1167}{{\tt arXiv:0802.1167}}].

\bibitem{Seery0610210}
D.~Seery, J.~E. Lidsey, and M.~S. Sloth, {\it {The inflationary trispectrum}},
  {\em JCAP} {\bf 0701} (2007) 027,
  [\href{http://xxx.lanl.gov/abs/astro-ph/0610210}{{\tt astro-ph/0610210}}].

\bibitem{Arroja09053641}
F.~Arroja, S.~Mizuno, K.~Koyama, and T.~Tanaka, {\it {On the full trispectrum
  in single field DBI-inflation}},  {\em Phys.Rev.} {\bf D80} (2009) 043527,
  [\href{http://xxx.lanl.gov/abs/0905.3641}{{\tt arXiv:0905.3641}}].

\bibitem{Chen09053494}
X.~Chen, B.~Hu, M.-x. Huang, G.~Shiu, and Y.~Wang, {\it {Large Primordial
  Trispectra in General Single Field Inflation}},  {\em JCAP} {\bf 0908} (2009)
  008, [\href{http://xxx.lanl.gov/abs/0905.3494}{{\tt arXiv:0905.3494}}].

\bibitem{Babich0405356}
D.~Babich, P.~Creminelli, and M.~Zaldarriaga, {\it {The Shape of
  non-Gaussianities}},  {\em JCAP} {\bf 0408} (2004) 009,
  [\href{http://xxx.lanl.gov/abs/astro-ph/0405356}{{\tt astro-ph/0405356}}].

\bibitem{Fergusson08123413}
J.~Fergusson and E.~Shellard, {\it {The shape of primordial non-Gaussianity and
  the CMB bispectrum}},  {\em Phys.Rev.} {\bf D80} (2009) 043510,
  [\href{http://xxx.lanl.gov/abs/0812.3413}{{\tt arXiv:0812.3413}}].

\bibitem{Komatsu09024759}
E.~Komatsu, N.~Afshordi, N.~Bartolo, D.~Baumann, J.~Bond, {\em et.~al.}, {\it
  {Non-Gaussianity as a Probe of the Physics of the Primordial Universe and the
  Astrophysics of the Low Redshift Universe}},
  \href{http://xxx.lanl.gov/abs/0902.4759}{{\tt arXiv:0902.4759}}.

\bibitem{Planck0604069}
{\bf Planck} Collaboration, {\it {The Scientific programme of planck}},
  \href{http://xxx.lanl.gov/abs/astro-ph/0604069}{{\tt astro-ph/0604069}}.

\bibitem{Komatsu10014538}
{\bf WMAP} Collaboration, E.~Komatsu {\em et.~al.}, {\it {Seven-Year Wilkinson
  Microwave Anisotropy Probe (WMAP) Observations: Cosmological
  Interpretation}},  {\em Astrophys.J.Suppl.} {\bf 192} (2011) 18,
  [\href{http://xxx.lanl.gov/abs/1001.4538}{{\tt arXiv:1001.4538}}].

\bibitem{Larsen0202127}
F.~Larsen, J.~P. van~der Schaar, and R.~G. Leigh, {\it {De Sitter holography
  and the cosmic microwave background}},  {\em JHEP} {\bf 0204} (2002) 047,
  [\href{http://xxx.lanl.gov/abs/hep-th/0202127}{{\tt hep-th/0202127}}].

\bibitem{vanderSchaar0307271}
J.~P. van~der Schaar, {\it {Inflationary perturbations from deformed CFT}},
  {\em JHEP} {\bf 0401} (2004) 070,
  [\href{http://xxx.lanl.gov/abs/hep-th/0307271}{{\tt hep-th/0307271}}].

\bibitem{Larsen0307026}
F.~Larsen and R.~McNees, {\it {Inflation and de Sitter holography}},  {\em
  JHEP} {\bf 07} (2003) 051,
  [\href{http://xxx.lanl.gov/abs/hep-th/0307026}{{\tt hep-th/0307026}}].

\bibitem{Assassi12044207}
V.~Assassi, D.~Baumann, and D.~Green, {\it {On Soft Limits of Inflationary
  Correlation Functions}},  \href{http://xxx.lanl.gov/abs/1204.4207}{{\tt
  arXiv:1204.4207}}.

\bibitem{Kehagias12051523}
A.~Kehagias and A.~Riotto, {\it {Operator Product Expansion of Inflationary
  Correlators and Conformal Symmetry of de Sitter}},
  \href{http://xxx.lanl.gov/abs/1205.1523}{{\tt arXiv:1205.1523}}.

\bibitem{Kehagias12101918}
A.~Kehagias and A.~Riotto, {\it {The Four-point Correlator in Multifield
  Inflation, the Operator Product Expansion and the Symmetries of de Sitter}},
  \href{http://xxx.lanl.gov/abs/1210.1918}{{\tt arXiv:1210.1918}}.

\bibitem{Assassi12107792}
V.~Assassi, D.~Baumann, and D.~Green, {\it {Symmetries and Loops in
  Inflation}},  \href{http://xxx.lanl.gov/abs/1210.7792}{{\tt
  arXiv:1210.7792}}.

\bibitem{Maldacena11042846}
J.~M. Maldacena and G.~L. Pimentel, {\it {On graviton non-Gaussianities during
  inflation}},  {\em JHEP} {\bf 1109} (2011) 045,
  [\href{http://xxx.lanl.gov/abs/1104.2846}{{\tt arXiv:1104.2846}}].

\bibitem{Antoniadis11034164}
I.~Antoniadis, P.~O. Mazur, and E.~Mottola, {\it {Conformal Invariance, Dark
  Energy, and CMB Non-Gaussianity}},
  \href{http://xxx.lanl.gov/abs/1103.4164}{{\tt arXiv:1103.4164}}.

\bibitem{Creminelli11080874}
P.~Creminelli, {\it {Conformal invariance of scalar perturbations in
  inflation}},  \href{http://xxx.lanl.gov/abs/1108.0874}{{\tt
  arXiv:1108.0874}}.

\bibitem{Hinterbichler11061428}
K.~Hinterbichler and J.~Khoury, {\it {The Pseudo-Conformal Universe: Scale
  Invariance from Spontaneous Breaking of Conformal Symmetry}},
  \href{http://xxx.lanl.gov/abs/1106.1428}{{\tt arXiv:1106.1428}}.

\bibitem{Hinterbichler12026056}
K.~Hinterbichler, A.~Joyce, and J.~Khoury, {\it {Non-linear Realizations of
  Conformal Symmetry and Effective Field Theory for the Pseudo-Conformal
  Universe}},  {\em JCAP} {\bf 1206} (2012) 043,
  [\href{http://xxx.lanl.gov/abs/1202.6056}{{\tt arXiv:1202.6056}}].

\bibitem{Creminelli12034595}
P.~Creminelli, J.~Norena, and M.~Simonovic, {\it {Conformal consistency
  relations for single-field inflation}},
  \href{http://xxx.lanl.gov/abs/1203.4595}{{\tt arXiv:1203.4595}}.

\bibitem{Hinterbichler12036351}
K.~Hinterbichler, L.~Hui, and J.~Khoury, {\it {Conformal Symmetries of
  Adiabatic Modes in Cosmology}},
  \href{http://xxx.lanl.gov/abs/1203.6351}{{\tt arXiv:1203.6351}}.

\bibitem{Seery0604209}
D.~Seery and J.~E. Lidsey, {\it {Non-Gaussian Inflationary Perturbations from
  the dS/CFT Correspondence}},  {\em JCAP} {\bf 0606} (2006) 001,
  [\href{http://xxx.lanl.gov/abs/astro-ph/0604209}{{\tt astro-ph/0604209}}].

\bibitem{McFadden10110452}
P.~McFadden and K.~Skenderis, {\it {Holographic Non-Gaussianity}},  {\em JCAP}
  {\bf 1105} (2011) 013, [\href{http://xxx.lanl.gov/abs/1011.0452}{{\tt
  arXiv:1011.0452}}].

\bibitem{McFadden11043894}
P.~McFadden and K.~Skenderis, {\it {Cosmological 3-point correlators from
  holography}},  {\em JCAP} {\bf 1106} (2011) 030,
  [\href{http://xxx.lanl.gov/abs/1104.3894}{{\tt arXiv:1104.3894}}].

\bibitem{Bzowski11121967}
A.~Bzowski, P.~McFadden, and K.~Skenderis, {\it {Holographic predictions for
  cosmological 3-point functions}},
  \href{http://xxx.lanl.gov/abs/1112.1967}{{\tt arXiv:1112.1967}}.

\bibitem{Coriano12100136}
C.~Coriano, L.~Delle~Rose, and M.~Serino, {\it {Three and Four Point Functions
  of Stress Energy Tensors in D=3 for the Analysis of Cosmological
  Non-Gaussianities}},  \href{http://xxx.lanl.gov/abs/1210.0136}{{\tt
  arXiv:1210.0136}}.

\bibitem{Strominger0106113}
A.~Strominger, {\it {The dS / CFT correspondence}},  {\em JHEP} {\bf 0110}
  (2001) 034, [\href{http://xxx.lanl.gov/abs/hep-th/0106113}{{\tt
  hep-th/0106113}}].

\bibitem{Strominger0110087}
A.~Strominger, {\it {Inflation and the dS / CFT correspondence}},  {\em JHEP}
  {\bf 0111} (2001) 049, [\href{http://xxx.lanl.gov/abs/hep-th/0110087}{{\tt
  hep-th/0110087}}].

\bibitem{Zamolodchikov87}
A.~B. Zamolodchikov, {\it {Renormalization Group and Perturbation Theory Near
  Fixed Points in Two-Dimensional Field Theory}},  {\em Soviets Journal of
  Nuclear Physics} {\bf 46} (1987) 1090.

\bibitem{Cardy:1989da}
J.~L. Cardy, {\it {Conformal invariance and statistical mechanics}},  {\em
  {Champs, Cordes et Ph\'enom\`enes Critiques, Les Houches, Session XLIX}}
  (1989).

\bibitem{Freedman0510126}
D.~Z. Freedman, M.~Headrick, and A.~Lawrence, {\it {On closed string tachyon
  dynamics}},  {\em Physical Review} {\bf D73} (2006) 066015,
  [\href{http://xxx.lanl.gov/abs/hep-th/0510126}{{\tt hep-th/0510126}}].

\bibitem{Bianchi0310129}
M.~Bianchi, M.~Prisco, and W.~M\"uck, {\it {New results on holographic three
  point functions}},  {\em JHEP} {\bf 0311} (2003) 052,
  [\href{http://xxx.lanl.gov/abs/hep-th/0310129}{{\tt hep-th/0310129}}].

\bibitem{Muck10062987}
W.~M\"uck, {\it {Running Scaling Dimensions in Holographic Renormalization
  Group Flows}},  {\em JHEP} {\bf 1008} (2010) 085,
  [\href{http://xxx.lanl.gov/abs/1006.2987}{{\tt arXiv:1006.2987}}].

\bibitem{DiFrancesco:1997nk}
P.~Di~Francesco, P.~Mathieu, and D.~Senechal, {\em {Conformal field theory}}.
\newblock Springer-Verlag, Inc., New York, 1997.

\bibitem{Creminelli0407059}
P.~Creminelli and M.~Zaldarriaga, {\it {Single field consistency relation for
  the 3-point function}},  {\em JCAP} {\bf 0410} (2004) 006,
  [\href{http://xxx.lanl.gov/abs/astro-ph/0407059}{{\tt astro-ph/0407059}}].

\bibitem{Cheung07090295}
C.~Cheung, A.~L. Fitzpatrick, J.~Kaplan, and L.~Senatore, {\it {On the
  consistency relation of the 3-point function in single field inflation}},
  {\em JCAP} {\bf 0802} (2008) 021,
  [\href{http://xxx.lanl.gov/abs/0709.0295}{{\tt arXiv:0709.0295}}].

\bibitem{Callan:1970yg}
J.~Callan, Curtis~G., {\it {Broken scale invariance in scalar field theory}},
  {\em Phys.Rev.} {\bf D2} (1970) 1541--1547.

\bibitem{Symanzik:1970rt}
K.~Symanzik, {\it {Small distance behavior in field theory and power
  counting}},  {\em Commun.Math.Phys.} {\bf 18} (1970) 227--246.

\bibitem{Symanzik:1971vw}
K.~Symanzik, {\it {Small distance behavior analysis and Wilson expansion}},
  {\em Commun.Math.Phys.} {\bf 23} (1971) 49--86.

\bibitem{CallanColemanJackiw70}
C.~G. Callan, Jr., S.~R. Coleman, and R.~Jackiw, {\it {A New improved energy -
  momentum tensor}},  {\em Ann. Phys.} {\bf 59} (1970) 42--73.

\bibitem{Underwood10094200}
B.~Underwood, {\it {A Breathing Mode for Warped Compactifications}},  {\em
  Class.Quant.Grav.} {\bf 28} (2011) 195013,
  [\href{http://xxx.lanl.gov/abs/1009.4200}{{\tt arXiv:1009.4200}}].

\bibitem{Bzowski:2012ih}
A.~Bzowski, P.~McFadden, and K.~Skenderis, {\it {Holography for inflation using
  conformal perturbation theory}},
  \href{http://xxx.lanl.gov/abs/1211.4550}{{\tt arXiv:1211.4550}}.

\bibitem{Mata:2012bx}
I.~Mata, S.~Raju, and S.~Trivedi, {\it {CMB from CFT}},
  \href{http://xxx.lanl.gov/abs/1211.5482}{{\tt arXiv:1211.5482}}.

\end{thebibliography}

\providecommand{\href}[2]{#2}\begingroup\raggedright\endgroup

\end{document}